%% file: imc_workshop_tex/paper.tex
\documentclass[10pt,sigconf,letterpaper,nonacm]{acmart}

\usepackage[english]{babel}
\usepackage{blindtext}
\usepackage{multirow}
\usepackage{dcolumn}
\usepackage[dvipsnames]{xcolor}
\usepackage{graphicx}
\usepackage{caption}
\usepackage{array}
\usepackage{cleveref}
\usepackage{subcaption} 
\usepackage{float}      
\usepackage{makecell}
\usepackage{balance}
\usepackage[inline]{enumitem}
\usepackage[most]{tcolorbox}

\renewcommand\footnotetextcopyrightpermission[1]{} 
\setcopyright{none}

\acmConference[Submitted for review to IMC Student Workshop]{}
\acmYear{2025}

\begin{document}
\title[]{The Potential of Erroneous Outbound Traffic Analysis to Unveil Silent Internal Anomalies}


\author{Andrea Sordello}
\orcid{0009-0005-9226-9887}
\affiliation{%
  \institution{Politecnico di Torino}
  \country{}
}

\author{Zhihao Wang}
\orcid{0000-0003-1989-3067}
\affiliation{%
  \institution{UESTC}
\country{}
}

\author{Kai Huang}
 \orcid{0009-0008-9272-5274}
\affiliation{%
  \institution{Politecnico di Torino}
\country{}
}

\author{Alessandro Cornacchia}
\orcid{0000-0002-4734-3321}
\affiliation{%
  \institution{KAUST}
\country{}
}

\author{Marco Mellia}
\orcid{0000-0003-1859-6693}
\affiliation{%
  \institution{Politecnico di Torino}
\country{}
}



\newif\ifsubmission
  \submissiontrue

\ifsubmission
    \newcommand{\kh}[1]{}
    \newcommand{\as}[1]{}
    \newcommand{\ac}[1]{}
    \newcommand{\mm}[1]{}
    \newcommand{\id}[1]{}
    \newcommand{\zw}[1]{}
\else
    \newcommand{\kh}[1]{\textit{\color{magenta}[Kai: #1]}}
    \newcommand{\as}[1]{\textit{\color{cyan}[Andrea: #1]}}
    \newcommand{\mm}[1]{\textit{\color{purple}[Marco: #1]}}
    \newcommand{\id}[1]{\textit{\color{red}[Idilio: #1]}}
    \newcommand{\zw}[1]{\textit{\color{orange}[Zhihao: #1]}}
    \newcommand{\ac}[1]{\textit{\color{ForestGreen}[Alessandro: #1]}}
\fi

\newcommand{\sysname}{\textsc{ACMETool}\xspace}
\newcommand{\smartparagraph}[1]{\noindent{\bf #1}.\ }

\newcounter{takeaway}
\newcommand{\takeawaycounter}{%
  \stepcounter{takeaway}%
  \arabic{takeaway}%
}
\newcounter{case}
\newcommand{\casecounter}{%
  \stepcounter{case}%
  Anomaly \arabic{case}: 
}

\definecolor{lightgraybox}{RGB}{245,245,245}
\definecolor{grayframe}{RGB}{180,180,180}

\newtcolorbox{takeawaybox}[1][]{
  colback=lightgraybox,
  colframe=grayframe,
  fonttitle=\bfseries,
  boxrule=0.5pt,
  sharp corners,
  left=3pt,
  right=3pt,
  top=3pt,
  bottom=3pt,
  title=#1,
}

\crefname{figure}{Fig.}{Figs.}
\Crefname{figure}{Fig.}{Figs.}

\crefname{tabular}{Table}{Tables}
\Crefname{tabular}{Table}{Tables}

\crefname{section}{Sec.}{Secs.}
\Crefname{section}{Sec.}{Secs.}

\newcommand{\circled}[1]{\tikz[baseline=(myanchor.base)] \node[circle,fill=.,inner sep=1pt] (myanchor) {\color{-.}\bfseries\footnotesize #1};}


\maketitle

\input{imc_workshop_tex/body_paper}

\bibliographystyle{ACM-Reference-Format}
\bibliography{reference}

\end{document}

%% file: imc_workshop_tex/body_paper.tex
\vspace{-1em}
\section{Research objective}
\vspace{-0.5em}
Network administrators have long relied on passive measurement to detect malicious activity, diagnose misconfigurations, and ensure network health.
Under the assumption that threats and issues originate externally, prior studies~\cite{GriffioenHaveYouSYNMe2024} have predominantly focused on the analysis of inbound traffic --- \emph{i.e.}, traffic initiated by external hosts and targeting internal destinations.
Conversely, outbound traffic --- \emph{i.e.}, originating from internal hosts toward external destinations --- has received comparatively less attention, despite carrying strong indicators of security-relevant anomalies~\cite{miao2015thedark}.

In this work, we offer a complementary perspective by focusing on this often-overlooked subset of traffic.
Unlike related works, we explore whether a narrow focus on a subset of outbound traffic that we term \emph{erroneous traffic} -- \emph{i.e.},  traffic that fails to complete a connection or triggers error responses -- would suffice to discover a broad class of malicious, suspicious, and anomalous patterns.

\begin{figure}[!t]
        \centering    
        \includegraphics[width=1\linewidth]{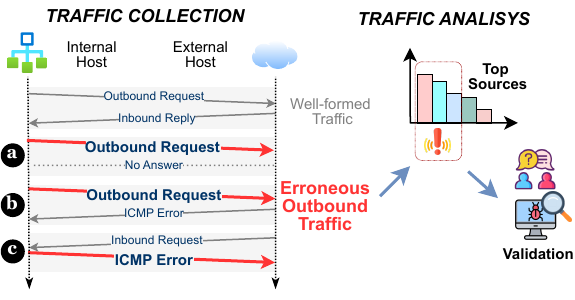}
        \caption{We study which anomalies can be uncovered by solely analysing \emph{erroneous} outbound traffic.}

\label{fig:outbound_erroneous_traffic}
\vspace{-1em}
\end{figure}

\smartparagraph{Erroneous outbound traffic and why it is valuable}
As shown in \cref{fig:outbound_erroneous_traffic}, \textit{erroneous outbound traffic} is made of packets sent by internal hosts toward external destinations that: \circled{a} elicit no response (e.g., offline/firewalled/inexistent destination), \circled{b} trigger an ICMP error (e.g., port/host/net unreachable), or \circled{c} carry ICMP errors generated by internal hosts themselves in response to incoming requests.
We term this traffic erroneous as in many cases it emerges from misconfigurations, faulty applications, or malicious activities. It indirectly reflects services that are operational but behave in unexpected or unintended ways, a scenario that can silently lead to security risks, wasted resources, low reliability, or a combination thereof.
To illustrate this, we conduct a preliminary analysis of erroneous outbound traffic collected from our campus to unveil several anomalies that have fallen under the radar for a long time, including compromised internal machines, faulty DNS clients, and decommissioned deployments, among others.
We believe erroneous outbound traffic is an appealing source for passive network troubleshooting. It is easy to isolate from normal bidirectional flows, resulting in a high signal-to-noise ratio and making anomalies more detectable with limited false positives. Additionally, when properly isolated from the bulk of well-formed traffic, it generates a low volume of data to log. 

\vspace{-1em}
\section{Traffic collection}\label{sec:collection}
\vspace{-0.5em}
\smartparagraph{Dynamic detection \& filtering of erroneous traffic}
We implement a prototype based on a stateful SDN-based network monitor~\cite{chamaleonet} that logs exclusively erroneous traffic, disregarding well-formed flows --- i.e., flows that do not exhibit any of the erroneous patterns. 
These patterns are relatively easy to detect via a timeout (pattern \circled{a}) and stateful packet inspection (pattern \circled{b} and \circled{c}).  
Our monitor leverages a network function to detect, at runtime, any flow that exhibits one of the erroneous patterns. 
The network function coordinates with an SDN control plane to install appropriate SDN flow rules that instruct network switches to forward only erroneous traffic to a packet logger. Thanks to this tool, and differently from an IDS, we can dynamically identify and opportunistically collect exclusively erroneous traffic at runtime, disregarding the sheer volume of benign traffic. 
For this work, we collected three months of erroneous outbound traffic from the \texttt{/16} campus subnet, from $1^{st}$ Dec.  2024 to $28^{th}$ Feb. 2025, with around 300\,$k$  
erroneous outbound packets per hour, about 0.06\% of the total outbound rate.

\smartparagraph{Ethics}
The IT and Privacy Officers have approved this work. We obfuscate internal IP addresses and truncate any layer 4 payload if present to avoid identifying devices and users. We shared our findings only with the IT security team, which provided us with the information of a few public internal host to contextualise our results.
\vspace{-1em}
\section{Traffic analysis \& preliminary insights}\label{sec:analysis}
\vspace{-0.5em}
We show the potential of erroneous traffic by \textit{i)} statistical macro analysis, and \textit{ii)} representative cases. 

\begin{figure}[tb!]
    \centering    
    \begin{subfigure}[t]{0.48\linewidth}
        \includegraphics[width=\linewidth]{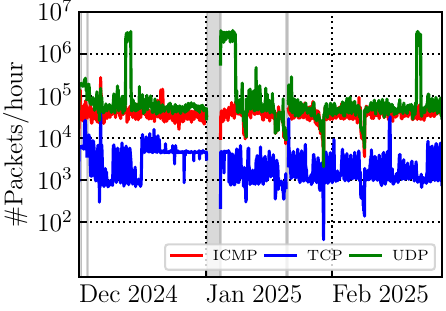}
        \caption{Persistent presence of erroneous packets}
        \label{fig:timeline-proto}
    \end{subfigure}
    \hfill
    \begin{subfigure}[t]{0.48\linewidth}
        \includegraphics[width=\linewidth]{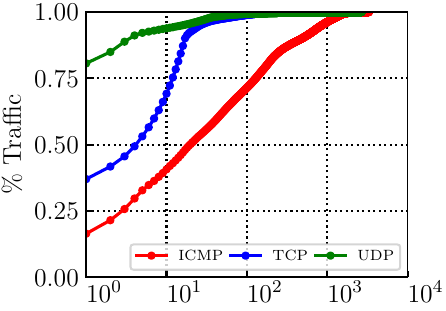}
        \caption{Few source IPs ($x$-axis) contribute disproportionately to overall erroneous traffic}
        \label{fig:cdf-senders}
    \end{subfigure}
    \vspace{-1em}
    \caption{Macro observations of erroneous traffic.}
    \label{fig:aggregate-analysis}
     \vspace{-1em}
\end{figure}

\smartparagraph{Macro observation}
\cref{fig:timeline-proto} presents the volume of erroneous outbound traffic, stably persisting over time. This suggests that erroneous traffic does not exclusively emerge from human activity but mainly from the continuous presence of misconfigured or malicious hosts. Spikes are related to a sudden rise in suspicious activities.
Second, the traffic distribution is highly skewed (\cref{fig:cdf-senders}), with a small subset of internal hosts responsible for the majority of the activity. 

\smartparagraph{Silent anomalies we discovered} We report a subset of representative and interesting anomalies, in terms of volume or periodicity, in \Cref{tab:erroneous-outbound}. These hosts generate 11\% of the total erroneous outbound traffic
. For each anomaly, we highlight the symptoms and the underlying root-cause, investigated and validated with our campus IT. None of which were previously known to our IT administrators.
We report the number of external ([E]) and internal ([I]) hosts involved and the number of packets ([P]) sent. 
We identify four categories:
\begin{enumerate*}[label=\arabic*)]
    \item \emph{Suspected malicious activity}; 
    We observe a sudden surge of ICMP port unreachable replies sent by about 700 internal hosts to CloudFlare 1.1.1.1 public DNS resolver, involving port 500/UDP. This could be a port scan that CloudFlare does for an unknown reason, or someone spoofing the 1.1.1.1 address for a reflection attack.
    %
    Next, we identify some internal hosts continuously attempting to contact IP addresses known to be associated with (dismissed) malicious infrastructure, indicating the presence of command-and-control (C2) channels or participation in a botnet.
    
    \item \textit{Host with faulty configuration}; hosts that generate erroneous outbound traffic due to faulty configuration, e.g., referencing non-active hostnames or using incorrect parameters. For instance, we observe more than 30 clients generating DNS queries directed to a ``localhost'' DNS resolver typically used by Tailscale VPN.
    The leakage of this traffic outside localhost is a clear indication of misconfiguration. 
    
    \item \textit{Host with stale configuration}; it involves stale or dismissed deployments resulting from, as an example, hosts with outdated configurations that try to reach no longer available repositories.
    
    \item \textit{Other}; 
    more than 50 internal hosts exhibiting unusual aggressive DNS lookups, 
    some of which we verified running DNS accelerators. These tools issue DNS requests to multiple open resolvers in parallel. The host accepts only the first DNS reply and closes the socket. Each late response triggers an ICMP port unreachable.  

\end{enumerate*}

\begin{table}[!t]
\resizebox{\linewidth}{!}{  
    \begin{tabular}{cp{3.5cm}cccp{4.5cm}}
    \toprule
    \textbf{} & \makecell{\textbf{Anomaly symptom}}   & \multicolumn{2}{c}{ \textbf{Hosts}} &\makecell{ \small \textbf{Packets}} & \makecell{\textbf{Root-cause}}   \\
    & & \textbf{[I]} & \textbf{[E]}& \textbf{[P]} &\\
    \midrule
    
    \multirow{8}{*}{\rotatebox{90}{\textbf{Malicious}}}
    &ICMP Port Unreachable replies to traffic from 1.1.1.1 for port 500 & $>$\,700 & 1 & $>$\,700  & Portscan from 1.1.1.1; or reflection attack with spoofed 1.1.1.1 source IP \\
    \cmidrule(lr){2-6}
    &Periodic probing to a single external IP address & 1 & 1 & 229\,$k$ & Internal host contacting a LockBit 4.0 ransomware node\\ 
    \cmidrule(lr){2-6}
     &Repeated unreplied requests to SMTP port & 1 & 382 & 205\,$k$ &Targets are known offering mail spam servers\\
     
    \midrule
    \multirow{5}{*}{ \rotatebox{90}{\textbf{Faulty}}} & DNS queries to a bogon IP address& $>$\,30 & 1 & 581\,$k$ &Misconfigured clients using private DNS resolvers \\
    \cmidrule(lr){2-6}
    &Unanswered NTP sync & 6 & 3 &72\,$k$ & Misconfigured NTP clients querying unreachable peers \\ 
    \midrule
    \multirow{3}{*}{ \rotatebox{90}{\textbf{Stale}}}&Persistent HTTP traffic to not responding external server & 1 & 1 & 1.8\,$M$ &Decommissioned software repository contacted every second by our HPC facility \\ 
    \midrule
    \multirow{5}{*}{ \rotatebox{90}{\textbf{Other}}}&Campus DNS resolvers with unanswered queries & 3 & $>$\,71\,$k$ & 63\,$M$ & Under investigation. Most to .cn and .ir domains \\
    \cmidrule(lr){2-6}
    &ICMP Port Unreachable triggered by legitimate DNS replies & $\sim$\,50 &100-1\,$k$ & 3.5\,$M$ &Aggressive DNS resolution using DNS accelerator tools \\
    \bottomrule
    \end{tabular}
}
 \vspace{-1em}
\caption{Silent internal anomalies we discovered.
}
\label{tab:erroneous-outbound}
 \vspace{-0.2em}
\end{table}

\vspace{-1em}
\section{Future work}\label{sec:future-work}
\vspace{-0.5em}
Our preliminary analysis has demonstrated the value of erroneous outbound traffic to gain visibility on a wide class of otherwise latent anomalies. 
However, elusive cases remain, for which the challenge lies in the lack of ground-truth data.
We outline future directions to streamline and automate the analysis workflow in an unsupervised manner. 
The following analysis pipeline is part of our current and future work, and tackles two problems: \emph{identifying} anomalous symptoms from the collected erroneous outbound traffic and \emph{explaining} their root causes. 
It comprises:
\begin{enumerate*}[label=\emph{\roman*)}]
\item representation learning, to learn a model of the erroneous interactions --  via NLP 
or GNNs
;
\item clustering, to aggregate the artifacts -- e.g., embeddings -- of representation learning and isolate anomalies;
\item pattern mining, ~
to extract interpretable rules that govern the clusters;
\item LLM-based AI agents, to validate hypotheses and explain the likely root-causes: given the output of pattern mining, they automate the otherwise manual and cumbersome tasks of inspecting hosts and services.
\end{enumerate*}

\vspace{-1em}
\section*{Acknowledgment}
\vspace{-0.4em}
This work was supported by project SERICS (PE00000014) under the MUR National Recovery and Resilience Plan funded by the European Union - NextGenerationEU. 

\vspace{-1em}

